\documentclass[preprint,showpacs,preprintnumbers,amsmath,amssymb,nofootinbib]{revtex4}
\usepackage{graphicx}
\usepackage{dcolumn}
\usepackage{bm}

\usepackage{epsfig}

\newcommand{\ba}{\begin{array}}

\newcommand{\ea}{\end{array}}

\newcommand{\bd}{\begin{displaymath}}

\newcommand{\ed}{\end{displaymath}}

\newcommand{\be}{\begin{equation}}

\newcommand{\ee}{\end{equation}}

\newcommand{\bea}{\begin{eqnarray}}

\newcommand{\eea}{\end{eqnarray}}

\def\barr{\begin{array}}

\def\earr{\end{array}}

\newcommand{\beq}{\begin{eqnarray}}

\newcommand{\eqn}{\end{eqnarray}}

\def\e{\epsilon}

\def\ep{\epsilon^\prime}

\def\lp{\lambda^\prime}
\def\lps{\lambda^{\prime *}}

\def\etal{ {\em et al.}}

\def\q2 {q^2}

\def\N10{\widetilde \chi_1^0}

\def\Cp1{\widetilde \chi_1^+}

\def\Cm1{\widetilde \chi_1^-}

\def\C1pm{\widetilde \chi_1^\pm}

\def \rp{R\hspace{-.6em}/\;\:}
\def \rpp{R\hspace{-0.4em}/\;\:}

\begin{document}
\begin{flushright}
OITS-754
\end{flushright}
\preprint{}
\title{Constraints on New Physics from $K \rightarrow \pi \nu \bar{\nu}$ }
\author{$^1$N.G. Deshpande\footnote{e-mail: desh@uoregon.edu},
$^1$Dilip Kumar Ghosh\footnote{e-mail: dghosh@physics.uoregon.edu}
and $^2$Xiao-Gang He\footnote{e-mail: hexg@phys.ntu.edu.tw. On
leave of absence from Department of Physics, National Taiwan
University, Taipei. Corresponding author.}}
\address{$^1$Institute of Theoretical Sciences, University of
Oregon,
OR 97403\\
$^2$Department of Physics, Peking University, Beijing, China}

\begin{abstract}

We study constraints on new physics from the recent measurement of
Br ($K^+ \to \pi^+ \nu \bar \nu)$ by the E787 and E949
Collaborations. In our analysis we consider two models of new
physics: $(a)$ extra down type singlet quark model (EDSQ) and
$(b)$ $R$-parity violating Minimal Supersymmetric Standard
Model(MSSM).
We find that $K^+\to \pi^+ \nu\bar \nu$ along with other processes
like $K_L \to \mu^+ \mu^-, \epsilon^\prime/\epsilon $ provide
useful bounds on the parameter $U_{sd}$, characterizing the
off-diagonal $Z-d-\bar s $ coupling of model $(a)$. The bounds on
the ${\rm Re}(U_{sd})$ from
$(K_L \to \mu^+\mu^-)_{\rm SD}$ and ${\rm Im}(U_{sd})$ from
$\epsilon^\prime/\epsilon$ are so tight that the branching ratio
of $K^+\to \pi^+ \nu \bar \nu $ can exceed the standard model value
by at most a factor of two.
For model b), we also obtain stringent bounds on certain
combinations of product of two $\lp_{ijk}$ couplings originating
from $L$ number violating operator $ L_i Q_j D^c_k $ using $K^+
\to \pi^+ \nu \bar \nu $ and $K_L \to \mu^+ \mu^-$ processes. Even
with the stringent constraints on $U_{sd}$ (in model $(a)$) and on
products of two $\rp$couplings ( model $(b)$), we find
that the branching ratio ${\cal B}(K_L \to \pi^0 \nu \bar \nu)$ 
can be substantially different in both the
above models from those predicted in the standard model.

\end{abstract}
 \pacs{PACS numbers: 12.15.-y, 12.60-.i, 14.70.-e ..}

\maketitle

\section{Introduction}
Recently the experiment E949 at Brookhaven National Laboratory has
detected an event for the rare decay $K^+\rightarrow \pi^+ \nu
\bar\nu$. Combining previously reported two events by the
experiment E787, a branching ratio $B(K^+\rightarrow \pi^+ \nu
\bar\nu) = (1.47^{+1.30}_{-0.89})\times 10^{-10}$\cite{Artamonov}
has been obtained. The central value of this branching ratio is
about twice that of the standard model (SM) prediction
$B(K^+\rightarrow \pi^+ \nu \bar\nu) = (7.2\pm 2.1) \times
10^{-11}$ \cite{new,Buchalla1995vs,burasN,smz,Battaglia2003in,bfrs}. The SM
prediction and the combined E787 and E949 results are consistent
with each other within one standard deviation. Future improvement
of the experimental sensitivity will verify whether any meaningful
difference emerges. If the experimental result converges to the
present central value, it would be an indication of new physics
beyond the SM. In extensions of the SM there are new sources for
flavor changing neutral currents which can affect the branching
ratio for $K^+\to \pi^+ \nu\bar \nu$ and reproduce the central
value obtained by E787 and
E949~\cite{smz,bfrs,susy,buras1,baren,jaas,other}.

In this work we analyze the effect of the new E787 and E949
results on two class of models beyond the SM. The first one is the
flavor changing neutral current (FCNC) mediated by Z boson in
extra down-type quark singlet (EDQS) model. These extra down type
singlet quarks appear naturally in each 27-plet fermion generation
of $E_6$ Grand Unification Theories (GUTs)
\cite{jlr,bdpw,rizzo,bbp}. The mixing of these singlet quarks with
the three SM down type quarks
induces tree-level flavor changing neutral currents (FCNC) by Z
exchange. These tree-level FCNC couplings can have significant
effects on various kaon decay processes including $K \to \pi
\nu\bar\nu$.

The second model, we consider is the $R$-parity violating minimal
supersymmetric standard model(MSSM).
The superpotential of the MSSM contains operators which violate
lepton $(L)$ and baryon $(B)$ numbers. The simultaneous presence
of both lepton and baryon number violating  operators leads to
rapid proton decay which contradicts the experimental bound on
proton life-time \cite{prot_life}. In order to keep the proton
life-time within the experimental limit one has to impose certain
additional symmetry in the model so that the baryon and lepton
number violating interactions vanish.  In most cases, a discrete
multiplicative symmetry called $R$-parity \cite{fayet} is imposed,
where $R =(-1)^{3B+L+2S}$, and $S$ is the spin of the particle.
Under this new symmetry, all baryon and lepton number violating
operators in the superpotential with mass dimension less or equal
to four vanish. This not only forbids rapid proton decay but also
predicts stable lightest supersymmetric particle (LSP), which
escapes the detection providing a unique signature of $R$-parity
conserving MSSM. However, this symmetry is {\em ad hoc} in nature,
with no strong theoretical arguments in support of it. There are
many other discrete symmetries such as baryon parity and lepton
parity, both of which can remove the unwanted operators from the
superpotential thus preventing rapid proton decay. Since, there is
no direct evidence supporting either $R$-parity conserving or
$\rp$MSSM, it is interesting to probe the
consequences of $\rp$model (in such a way that the
either $B$ or $L$ number is violated but not both ) in light of
some recent low energy data. Already extensive studies have been
done to look for the direct as well as indirect evidence of
$R$-parity violation from different processes and to put
constraints on various $\rp$couplings \cite{rpv1,
rpv2,proton, Zdk,dc1,gamma,btolep1,gg1}. We consider
constraints on the product of two $\rp$couplings that ensue
from $K^+ \to \pi^+ \nu \bar \nu$ decay as
well as other kaon processes, such as $K_L \to \mu^+ \mu^-,
K^+ \to \pi^+ \mu^+ e^- $. In some cases, the bounds obtained in this
analysis are stronger than the existing one. Interestingly even with such
a stronger
bounds, for some combinations of $\rp$couplings, the
${\rm Br}(K_L \to \pi^0 \nu \bar \nu)$ 
is well above the standard model prediction.

The rest of the paper is organized in the following way. In
Section II we discuss the tree level FCNC effects in the extra
down type singlet quark model. We will then constrain this
new FCNC parameter using the latest data on ${\rm Br} (K^+ \to
\pi^+ \nu\bar\nu)$, ${\rm Br}(K_L \to \mu^+\mu^-)_{\rm SD}$, and
$\ep/\e$ in kaon decay into two pions. After constraining the FCNC
parameter space we look for the prediction for the $CP$ violating
processes: $K_L \to \pi^0 \nu \bar\nu$ and $K_L \to \pi^0 e^+ e^-$
in the allowed range of parameter space. In Section III we obtain
constraints on the product of two $\rp$couplings using the
current data on ${\rm Br} (K^+ \to \pi^+ \nu\bar\nu)$, and
${\rm Br}(K_L \to \mu^+\mu^-)_{\rm SD}$, followed by the prediction of
 $K_L \to \pi^0 \nu \bar \nu$ and process in the allowed range of the 
$\rp$couplings. We summarize our results in the last section.

\section{Z mediated FCNC with extra down type singlet quark}

FCNC mediated by Z boson contributing to $K^+ \to \pi^+ \nu \bar
\nu$ can be generated in many ways when going beyond the SM. A
simple possibility arises from one new vector-like down type
singlet quark in addition to the SM fermions. In the weak
interaction basis, W and Z interactions with quarks (in the
current basis) can be written as
\begin{eqnarray}
{\cal L}_W &=& -{g\over \sqrt{2}} \bar U^0_L \gamma^\mu D^0_L W^+,\nonumber\\
{\cal L}_Z &=& -{g\over 2 c_W}[\bar U^0_L\gamma^\mu U^0_L - \bar D^0_L
\gamma^\mu D^0_L - 2 s^2_W (Q_u\bar U^0\gamma^\mu U^0 + Q_d \bar D^0
\gamma^\mu D^0 + Q_d \bar D^{0\prime}\gamma^\mu D^{0 \prime})] Z_\mu,
\end{eqnarray}
where $U^0 = (u,c,t)$, and $D^0 = (d,s,b)$ are the usual three
generations of quarks in the SM, and $D^{0\prime} =d^{0\prime} $
is the additional
down type of quark singlet. One can easily generalize the model to
include $n$ generations of vector-like down-type quarks by using
$D^{0\prime} = (d_1^{0 \prime},...d^{0\prime}_n)$.

In general $D^{0\prime}$ can mix with the ordinary quarks in
$D^0$. The down quark mass matrix $M_d$ is diagonalized by
$4\times 4$ unitary matrices $V^{L\dagger}_d M_d V^{R}_d =
dig(m_d, m_s,m_b,m_{d'})$. While the up quark mass matrix $M_u$ is
diagonalized by $3\times 3$ unitary matrices $V^{L\dagger}_u M_u
V^R_u = dig(m_u, m_c, m_t)$. We indicate mass diagonal basis
$D\equiv (d,s,b,d^\prime) $ and in this basis, we have
\begin{eqnarray}
{\cal L}_W &=& -{g\over \sqrt{2}} \bar U_L \gamma^\mu V D_L W^+,\nonumber\\
{\cal L}_Z &=& -{g\over 2 c_W}[\bar U_L\gamma^\mu U_L - \bar
D_L\gamma^\mu D_L - 2 s^2_W (Q_u\bar U\gamma^\mu U + Q_d \bar D
\gamma^\mu D ]Z_{\mu}\nonumber\\
& -& {g\over 2 c_W} \bar D_{Li}\gamma^\mu U_{ij}D_{Lj}Z_{\mu},
\label{intz}
\end{eqnarray}
where $Q_{u,d}$ are the electric charges of up and down quarks in
unit of proton charge. $U_{ij} = V^{L*}_{dil}V^L_{djl}$ is
$4\times 4$ matrix, and $V  =
V_u^{L\dagger}V^{L}_d$ which is a $3\times 4$ matrix different
from the usual KM matrix $V_{KM}$. We shall be concerned with only
$U_{sd}$ element of the $U$ matrix in this paper.

In the absence of $d'$, the theory reduces to the SM. The top-left
$3\times 3$ block matrix in $V$ corresponds to the usual $V_{KM}$
matrix. The rest of the matrix elements in $V$ are expected to be
small since deviations away from the SM are constrained to be
small from various experimental data.

\subsection{ $K^+\to \pi^+ \nu\bar \nu$ process }

\begin{figure}[hbt]
\centerline{\epsfig{file=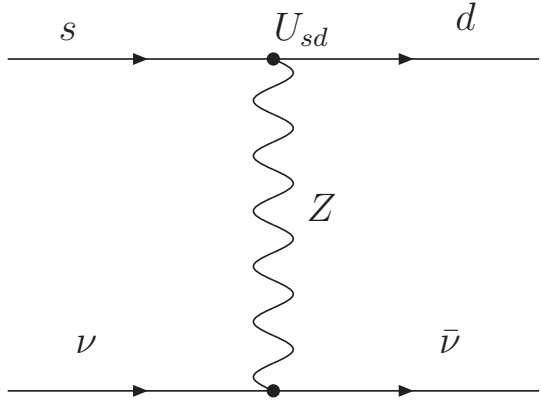,width=\linewidth}}
\vspace*{-16.0cm}
\caption{Feynman diagram for $Z$ exchange tree-level contribution to
$K^+ \to \pi^+ \nu \bar \nu$ process}
\label{diag}

\end{figure}

The FCNC interactions as described by ${\cal L}_{Z}$ in Equation \ref{intz}
can contribute to $K^+\to \pi^+ \nu\bar
\nu$ decay at the tree level by exchanging Z boson as shown in
Figure \ref{diag}. The relevant Lagrangian is given by
\begin{eqnarray}
{\cal L}_Z^{\nu} = {g^2\over 4 c^2_W m^2_Z} U_{sd}\bar s_L \gamma^\mu d_L
\sum_{\ell = e,\mu,\tau} \bar \nu_{L_\ell}\gamma_\mu  \nu_{L_\ell}.
\label{eff}
\end{eqnarray}

Using $<0|\bar s \gamma^\mu \gamma_5 d |K^+> = i f_K p^\mu_K$ and
including the SM contribution,  one obtains \cite{jaas}
\begin{eqnarray}
&&{{\rm Br} (K^+\to \pi^+ \nu \bar \nu)\over {\rm Br} (K^+ \to \pi^0 e^+\nu)}
={r_{K^+}\alpha^2 \over 2 \pi^2 s^4_W |V_{us}|^2}
 \sum_{\ell=e,\mu,\tau}|\Delta^{SM}_{K} + \Delta^Z_{K}|^2,
\nonumber\\
 &&\nonumber\\
&&\Delta^{SM}_{K} =\lambda^c_{sd} X^\ell_{NL} +
\lambda^t_{sd}\eta_t^X X_0(x_t)
\label{kpi_z}
\end{eqnarray}
The factor $r_{K^+}=0.901$ accounts for isospin breaking
corrections \cite{maciano}, $\alpha = 1/128 $, $s_W = \sin\theta_W$.
$\Delta^{SM}_{K}$ is the SM contribution with the
charm contributions at NLO found to be $X^{e,\mu}_{NL} =(10.6\pm
1.5)\times 10^{-4}$, $X^\tau_{NL} = (7.1\pm 1.4)\times 10^{-4}$ \cite{burasN}.
$\lambda^i_{sd}$ is defined as $V_{is}^*V_{id}$. The top
contribution is proportional to the term with $\lambda^t_{sd}$.
$\eta^X_t$ is a QCD correction factor which is equal to $0.994$.
The function $X_0$ is given by: 
\begin{eqnarray}
X_0 = {x\over 8} \left[ {x+2\over x-1} + {3x-6\over (x-1)^2} \ln x \right ];
~~~~x = \frac{m_t^2}{m_W^2}
\end{eqnarray}
Using the current best fit values for the Wolfenstein parameters,
$\lambda =0.224$, $A=0.839$, $\rho=0.178$, $\eta = 0.341$, and
experimental value of ${\rm Br} (K^+\to \pi^0 e^+ \nu) = 0.0487$, we
obtain the SM value of $(7.28^{+0.13}_{-0.12}) \times 10^{-11}$ for
${\rm Br} (K^+\to \pi^+ \nu \bar \nu)$. The theoretical error on the branching
ratio is computed by allowing $\lambda^t_{sd}$ and the charm NLO corrections
($X^{e,\mu}_{NL},X^\tau_{NL} $) to vary within $1\sigma$ from their central
value.  The central value of the SM branching ratio
is a factor of 2 smaller than the experimental central value.

The term $\Delta^Z_{K}$ in Equation (\ref{kpi_z}) characterizes
new contribution from FCNC $Z$ interaction which is given by:
\begin{eqnarray}
\Delta^Z_{K} = -{\pi s^2_W\over \alpha} U_{sd}.
\end{eqnarray}
With this new contributions it is
possible to reproduce the experimental central value, for example
with ${\rm Re} (U_{sd}) = 0.339 \times 10^{-5}$ and
${\rm Im } (U_{sd}) = 0.055 \times 10^{-5}$, we obtain
${\rm Br} (K^+\to \pi^+\nu\bar \nu)=  1.47 \times 10^{-10}$ using
the central values of $\lambda^c_{sd}, \lambda^t_{sd}$ and the charm NLO
corrections.

\begin{figure}[hbt]
\centerline{\epsfig{file=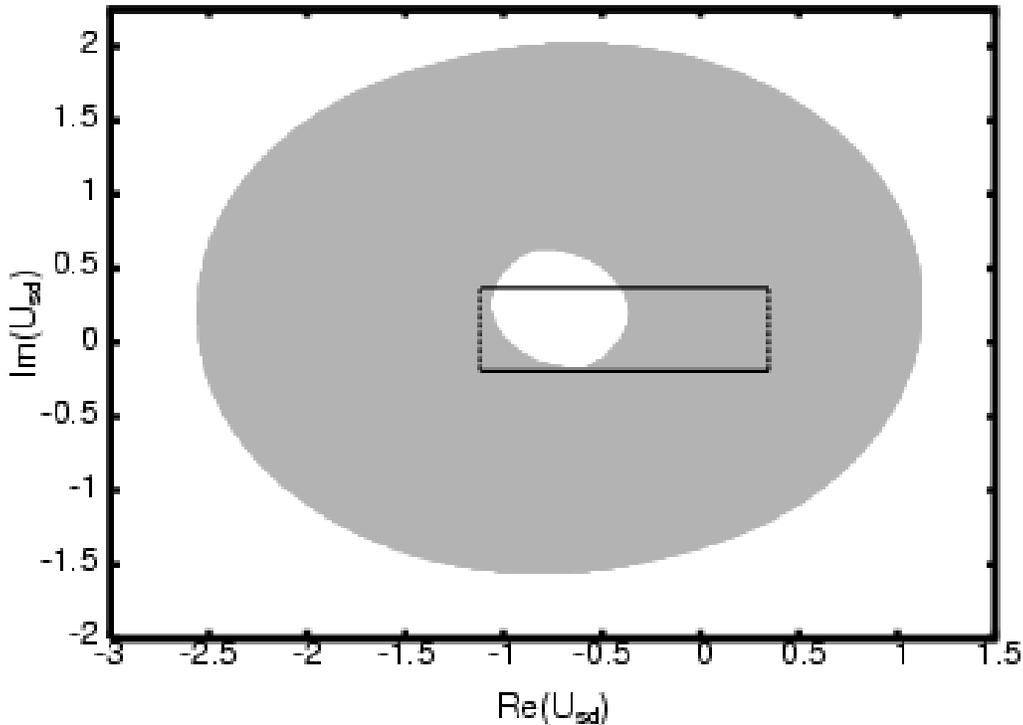,width=\linewidth}}
\vspace*{-10.0cm}
\caption{The shaded region is $90\%$ CL allowed parameter space
in ${\rm Re} (U_{sd})- {\rm Im} (U_{sd})$ plane from
${\rm Br}~(K^+ \to \pi^+ \nu \bar\nu)$ data.
The area enclosed by the two solid horizontal and two vertical dotted
lines is allowed by ${\rm Br}~(K_L \to \mu^+\mu^-)_{\rm SD}$
and $\epsilon^\prime/\epsilon$ respectively at $90\%$ CL. Both
${\rm Re} (U_{sd})$ and $ {\rm Im} (U_{sd})$ are measured in units
of $ 10^{-5}$.}
\label{kpi_allowed}

\end{figure}

One can also turn the argument around. By using the experimental
data on ${\rm Br }(K^+ \to \pi^+ \nu \bar \nu)$ one can constrain
the new FCNC parameter $U_{sd}$. In Figure \ref{kpi_allowed} the
grey shaded region represents $90 \%~{\rm CL} $ allowed region in
the ${\rm Re}~(U_{sd})-{\rm Im}~(U_{sd})$ plane from
${\rm Br}^{\rm exp}(K^+ \to \pi^+ \nu \bar \nu)$.
The thickness of the band includes the $1\sigma $ theoretical error
arising from the CKM parameters and the charm NLO contribution.
From the Figure \ref{kpi_allowed} one can see that the recent data
on ${\rm Br}(K^+ \to \pi^+\nu \bar \nu)$ still allows a large part
of the parameter space of $U_{sd}$. However, in next two
sub-sections we will show that the current information on the
short distance contribution to $K_L \to \mu^+\mu^- $ and the data
on $\ep/\e $ constrain the above allowed parameter space of
$U_{sd}$ quite severely.

\subsection{Constraints from other kaon processes}

The same flavor changing $Z$ interaction in Equation (\ref{intz})
will also have effects on other flavor changing kaon processes. The
$\Delta S = 1$
and $\Delta S = 2$ effective Lagrangian relevant for
our discussions are given
by
\begin{eqnarray}
&&{\cal L}^{\Delta S = 1} = {g^2\over 4 c_W^2 m^2_Z} U_{sd} \bar s_L
\gamma^\mu d_L \bar f \gamma_\mu (2I_3 L - 2Q_f s^2_W)
f,\nonumber\\
&&{\cal L}^{\Delta S = 2} = {G_F\over \sqrt{2}}
\left [ U_{sd}^2 -8{\alpha \over 4 \pi
 s^2_W}U_{sd}\sum_{\alpha=c,t}\lambda^\alpha_{sd}Y_0(x_\alpha) \right ]
\bar s_L \gamma^\mu d_L\bar s_L \gamma_\mu d_L
\end{eqnarray}
where, $Y_0(x) = (x/8)\left[ (x-4)/(x-1)+ 3 x \log x/(x-1)^2
\right ]$. The second term in $L^{\Delta S=2}$ is due to one loop
effect which is important for gauge invariance, but is numerically
small \cite{jaas}. Taking $f = \nu, e, \mu$, $L^{\Delta S = 1}$
can contribute to processes like $K^+\to \pi^+ \bar \nu \nu$, $K_L
\to \pi^0 \nu\bar\nu$, $K_L \to \mu^+\mu^-$ and $K_L\to \pi^0
e^+e^-$. Taking $f=u,d$, $L^{\Delta S=1}$ can contribute to
$\epsilon'/\epsilon$. While $L^{\Delta S=2}$ can contribute to
$\epsilon_K$ and $\Delta M_K$.
\\

\noindent {\bf Constraint from $K_L \to \mu^+\mu^-$}

An interesting limit on the FCNC parameter $U_{sd}$ comes from the
short-distance contribution to the decay $K_L \to \mu^+\mu^- $.
The $K_L \to \mu^+\mu^-$ branching ratio can be decomposed into
dispersive part (${\rm Re}~A)$ and the absorptive part (${\rm
Im}~A )$. The absorptive can be determined very accurately from
the branching ratio ${\rm Br}(K_L \to \gamma \gamma)$ and the
resulting $\mid {\rm Im}~A \mid^2 $ alone almost saturates the
branching ratio ${\rm Br} (K_L \to \mu^+ \mu^-) = (7.07 \pm
0.18)\times 10^{-9}$ \cite{jaas}
leaving a very small room for the dispersive
contribution ${\rm Re}~A $. ${\rm Re}~A $ can be further
decomposed in ${\rm Re}~A_{LD} + {\rm Re }~A_{SD}$.
Combining results from Ref.\cite{bfrs,isidori1},
we have $90\%$CL bound on ${\rm Br}(K_L \to \mu^+ \mu^-)_{SD}\leq 2.5
\times 10^{-9}$, with $\mid {\rm Re}~A_{SD}\mid^2
\equiv {\rm Br}(K_L \to \mu^+ \mu^-)_{SD}$. In our analysis
we use this value of ${\rm Br}(K_L \to \mu^+ \mu^-)_{SD}$ to constrain
the FCNC parameter ${\rm Re} (U_{sd}) $. The expression for
${\rm Br} (K_L \to \mu^+\mu^-)){\rm SD}$ is given by \cite{jaas}:
\begin{eqnarray}
&&\frac{ {\rm Br} (K_L \to \mu^+ \mu^-)_{\rm SD}}{ {\rm Br}(K^+
\to \mu^+ \nu)} = \frac{\tau_{K_L}}{\tau_{K^+}}\frac{\alpha^2}
{\pi^2 s^4_W \mid V_{us}\mid^2} \times \left[ T_{SM} + {\rm Re}
(\Delta^Z_{K})\right ]^2, \label{klmm}
\end{eqnarray}
where,
\beq T_{SM} &=& Y_{NL} {\rm Re}~(\lambda^c_{sd}) + \eta^Y_t
Y_0(x_t)
{\rm Re}~(\lambda^t_{sd})
\label{klmm1}
\eqn
 At $90\%$CL we get following bound on ${\rm Re} (U_{sd})$:
\beq
-1.12 \times 10^{-5} \leq {\rm Re} (U_{sd}) \leq 3.45 \times
10^{-6}
\eqn
The lower bound is obtained by taking
${\rm Re}(\lambda^c_{sd}) = (-0.2204 -0.0023)$, ${\rm
Re}(\lambda^t_{sd}) = (-3.04 - 0.31)\times 10^{-4}$ and the charm
NLO contribution $Y_{NL} = (2.94 +0.28) \times 10^{-4}$. The upper
bound is obtained by using ${\rm Re}(\lambda^c_{sd}) = (-0.2204
+0.0022)$, ${\rm Re}(\lambda^t_{sd}) = (-3.04 +0.32)\times
10^{-4}$ and the charm NLO contribution $Y_{NL} = (2.94 -0.28)
\times 10^{-4}$. In Figure \ref{kpi_allowed} the above bound is
shown by the area enclosed by vertical dotted lines. This shows that
a substantial part of the parameter space
which was allowed by the ${\rm Br}(K^+ \to \pi^+\nu \bar\nu)$ is
ruled out by the ${\rm Br}(K_L \to \mu^+\mu^-)_{\rm SD}$.

\noindent {\bf Constraint from $\epsilon^\prime /\epsilon $}

Following the notation of Ref.\cite{jaas}, we write the down the
expression for $\epsilon^\prime/\epsilon$:
\beq \ep/\e = F_{\ep}(x_t) {\rm Im} (\lambda^t_{sd}) +
\Delta_{\ep}
\eqn
where, $F_{\ep}(x_t)$ and can be found in the
Ref.\cite{jaas} and $ \Delta_{\ep} = -\frac{\pi
s^2_W}{\alpha} (P_X + P_Y + P_Z) {\rm Im} (U_{sd})$, with
$P_{X,Y,Z}$ are given in Ref.\cite{jaas}.

The bound on ${\rm Im} (U_{sd})$ from $\ep/\e$ depends upon the
sign of ${\rm Im} (U_{sd})$. For ${\rm Im}(U_{sd})> 0 $, the upper
bound looks like :
\beq
{\rm Im} (U_{sd}) \leq \frac{(\ep/\e)^{\rm
exp} - (\ep/\e)^{\rm SM} }{-\frac{\pi s^2_W}{\alpha} (P_X
+ P_Y + P_Z)}
\eqn

The experimental value for $\ep/\e$ is $(1.8
\pm 0.4)\times 10^{-3}$. To maximize the ${\rm Im} (U_{sd})$, one
should take maximum and minimum values for $(\ep/\e)^{\rm exp}$
and $(\ep/\e)^{\rm SM}$ respectively. The minimum value for the
$(\ep/\e)^{\rm SM}$ can be obtained by taking the lowest allowed
values of ${\rm Im} (\lambda^t_{sd})$, $B_6$ and $B_8$ (defined in
Ref.\cite{jaas}). With these choices of input parameters we have
found
\beq {\rm Im } (U_{sd}) \leq 3.72 \times 10^{-6} \label{eps1}
\eqn

Now, for ${\rm Im} (U_{sd}) < 0 $, we have
\beq
-{\rm Im} (U_{sd}) \leq \frac{(\ep/\e)^{\rm SM} - (\ep/\e)^{\rm exp} }
{-\frac{\pi s^2_W}{\alpha} (P_X + P_Y + P_Z)}
\eqn
In this case we take $(\ep/\e)^{\rm SM}_{\rm max} $ and
$(\ep/\e)^{\rm exp}_{\rm min.}$. This can be achieved by taking
largest allowed values of ${\rm Im} (\lambda^t_{sd})$,
$B_6$ and $B_8$. This gives us
\beq
-{\rm Im} (U_{sd}) \leq 1.91 \times 10^{-6}
\label{eps2}
\eqn

Clearly the bounds on the FCNC parameter ${\rm Im} (U_{sd})$ depends upon the
experimental values of $(\ep/\e)_{\rm max}$ and $(\ep/\e)_{\rm min}$.
Finally combining Equation \ref{eps1} and Equation \ref{eps2} we get a
bound on ${\rm Im} (U_{sd})$ at $ 90\% $CL :
\beq
-1.91 \times 10^{-6} \leq {\rm Im} (U_{sd}) \leq 3.72 \times 10^{-6}
\eqn
In Figure \ref{kpi_allowed} this bound is depicted by area enclosed
by two parallel solid lines. This bound further constrain the
parameter space of $U_{sd}$ which was otherwise allowed by the
${\rm Br}(K^+ \to \pi^+ \nu \bar \nu)$ and
${\rm Br}(K_L \to \mu^+\mu^-)_{\rm SD}$.

\noindent {\bf Remark on $\Delta M_K$ and $\epsilon_K$}

 The contribution from the new Z flavor changing neutral current
to $\Delta M_K^Z$ and $\epsilon_K^Z$ are given by
\begin{eqnarray}
M^K_{12} &=& {G_F^2 M^2_W f^2_K B_K M_K\over 12 \pi^2} [-8U_{ds}(
\lambda^c_{ds}\eta_c Y_0(x_c) + \lambda_{ds}^t \eta_t Y_0(x_t))
+{4\pi s^2_W\over \alpha} \eta_c U^2_{sd}],\nonumber\\
 \Delta M_K^{Z} &=& 2{\rm Re}
(M^K_{12}),\;\;\epsilon_K^Z = e^{i\pi/4}
{{\rm Im} (M^K_{12})\over \sqrt{2}\Delta M_K}.
\end{eqnarray}

We observe that $\Delta M_K $ and $\epsilon_K$ evaluated
from the above are
close to the standard model predictions after constraints from $K^+
\to \pi^+ \nu \bar \nu$ and $K_L\to \mu^+\mu^-$ is taken
into account. Uncertainties in long distance contributions to
$\Delta M_{K}$ make these process unsuitable for obtaining strong
constraints.

\begin{figure}[hbt]
\centerline{\epsfig{file=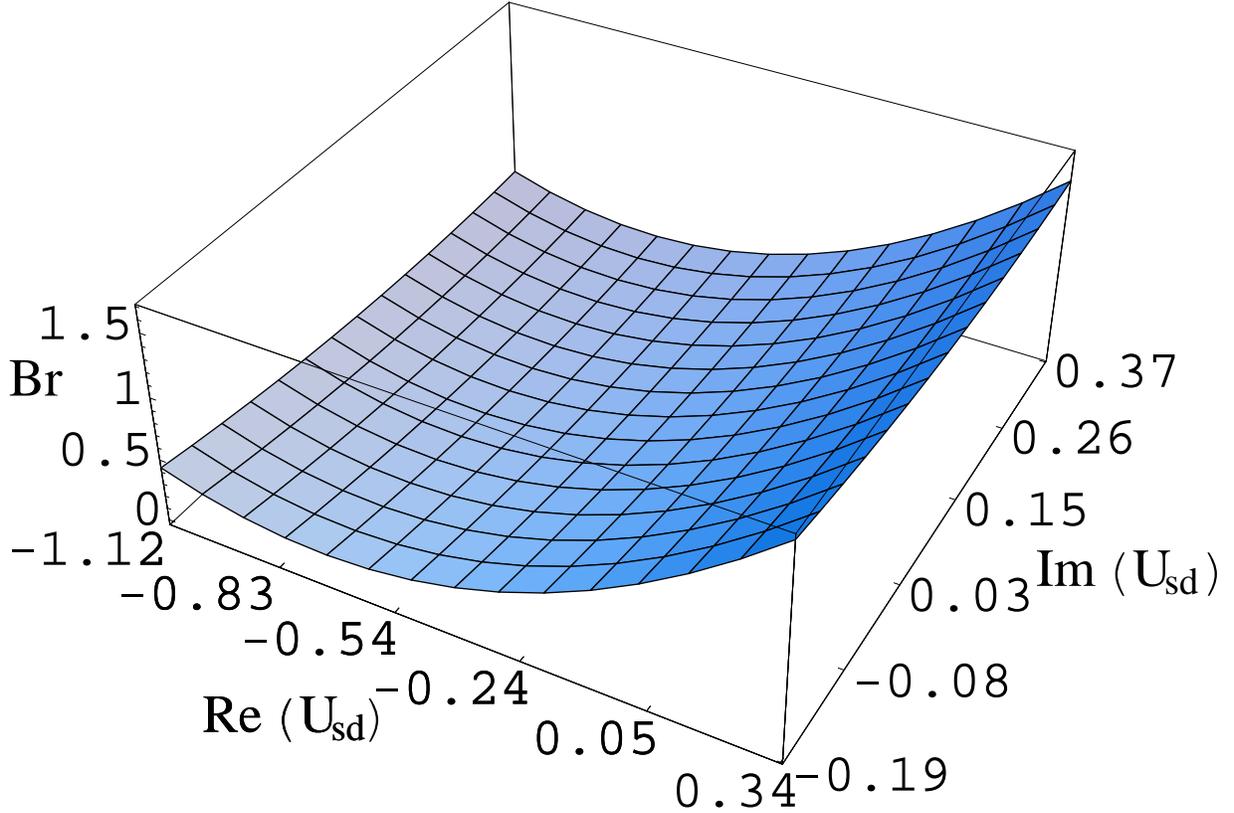,width=\linewidth}}
\caption{Variation of ${\rm Br}~(K^+ \to \pi^+ \nu \bar\nu)$
(in units of $10^{-10}$) in the allowed parameter space of the
${\rm Re}~(U_{sd})$ and ${\rm Im}~(U_{sd})$~(both in units of $10^{-5})$
as shown by the rectangular box in Figure \ref{kpi_allowed}.}

\label{3df}

\end{figure}

\noindent {\bf Predictions for $K^+ \to \pi^+ \nu \bar \nu$,
$ K_L \to \pi^0 \nu \bar \nu$ and $K_L\to \pi^0 e^+ e^-$}

\begin{figure}[hbt]
\centerline{\epsfig{file= 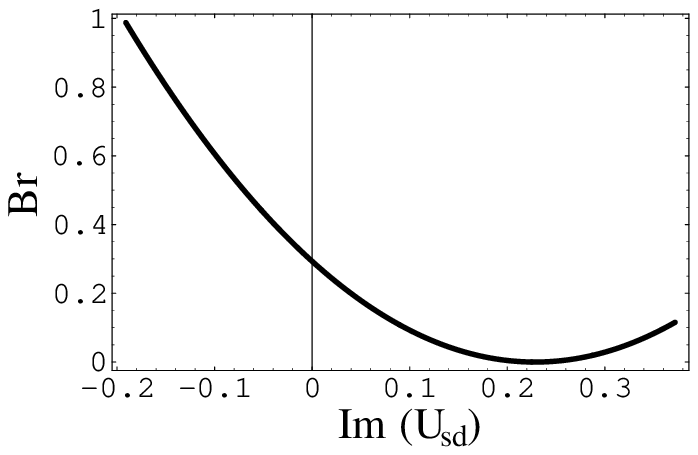,width=\linewidth}}
\caption{Variation of ${\rm Br}~(K_L \to \pi^0 \nu \bar\nu)$
(in units of $10^{-10}$) in the allowed range of ${\rm Im}~(U_{sd})
$~(in units of $10^{-5})$.}

\label{fig3}

\end{figure}

 In the previous discussions we have shown how recent data
on ${\rm Br} (K^+ \to \pi^+ \nu \bar \nu)$, ${\rm Br} (K_L \to \mu^+
\mu^-)_{\rm SD}$ and $\ep/\e$ could be used to constrain the tree
level FCNC parameter $U_{sd}$ in extra down type singlet quark
model. In view of the tight constraints on $U_{sd}$ obtained, we
now examine the range of allowed branching ratios for
$K^+ \to \pi^+ \nu \bar \nu$, $K_L \to
\pi^0 \nu \bar \nu$ and $K_L \to \pi^0 e^+ e^- $ within the
allowed parameter space of $U_{sd}$.

In Figure \ref{3df} we show the variation of ${\rm Br}(  K^+ \to
\pi^+ \nu \bar \nu) $ (in units of $10^{-10}$) in the allowed
${\rm Re} (U_{sd}) - {\rm Im} (U_{sd}) $ plane. It turns out that
the branching ratio can reach up to $1.5 \times 10^{-10}$ at the
edge of the allowed parameter ${\rm Re}~(U_{ds})$ and ${\rm Im}~(U_{ds})$.
Another point to note is that the branching ratio
depends very weakly on the FCNC parameter ${\rm Im}~(U_{ds})$. In
the computation of this branching ratio we consider only
the central values for all theoretical input parameters.

The channel $K_L \to \pi^0 \nu \bar \nu $ in the SM and in the
model under consideration is dominantly $CP$ violating. In the
standard model this decay mode is mostly determined by the
intermediate top quark state and the uncertainty due to the charm
contribution is negligible compared to $K^+ \to \pi^+ \nu \bar
\nu$ channel. We have \cite{jaas}
\begin{eqnarray}
&&{{\rm Br} (K_L\to \pi^0 \nu \bar \nu)\over {\rm Br} (K^+ \to \pi^0 e^+\nu)}
=r_{K_L}{\tau_{K_L}\over \tau_{K^+}}{\alpha^2 \over 2 \pi^2
s^4_W |V_{us}|^2}
 \sum_{\ell=e,\mu,\tau}|{\rm Im}~(\Delta^{SM}_{K} + \Delta^Z_{K})|^2.
\end{eqnarray}
where, $\Delta^{SM}_{K}$ and $\Delta^Z_{K}$ are defined in
Equation (4) and (6) respectively. The standard model branching fraction
for the process
$K_L \to \pi^0 \nu \bar\nu$ is $2.93^{+0.84}_{-0.67} \times 10^{-11}$ in the
same ballpark as other estimates \cite{bfrs,gisidori,kettell}.
The errors correspond to the $1\sigma $ error in the CKM elements.
In Figure \ref{fig3}, we show the variation of ${\rm Br} (K_L
\to \pi^0 \nu \bar \nu)$ (in units of $10^{-10}$) in the allowed
${\rm Re} (U_{sd}) - {\rm Im} (U_{sd})$ parameter space. Being a
$CP$ violating process, the branching ratio is solely dependent
upon the ${\rm Im}~(U_{sd})$ part of the FCNC parameter $U_{sd}$
and it can reach as high as $10 \times 10^{-11}$ at the edge of
the allowed region. Presently there is an upper bound on the
branching ratio of $5.9 \times 10^{-7}$ at $90\%$ CL \cite{prot_life}.

We now discuss the CP violating mode $K_L \to \pi^0 e^+ e^- $. It has
recently been shown \cite{buchalla_1, ishidori_1, friot_1} that there are
two main contributions to this process. The amplitude arising from the short
distance physics, and a mixing contribution arising from conversion of $K_L $
to $K_S$ and subsequent decay of $K_S$ into $\pi^0 e^+ e^-$. Using the 
experimental input on ${\rm Br} (K_S \to \pi^0 e^+ e^-)$, it has been shown
that this mixing contribution dominates over the short distance contribution 
in the rate by a factor of $5\sim 6$. As a result of this the new physics 
contribution to the short distance amplitude through $Z$ exchange, using 
constraints on the parameter $U_{sd}$ in Figure 2, hardly affects the total 
rate. The experimental bound on $K_L \to \pi^0 e^+e^-$ is larger than theory
prediction by an order of magnitude, so no stronger bounds on $U_{sd}$ can
be obtained.



\section{FCNC from $R$ parity violating interactions in MSSM}

The most general superpotential of the minimal supersymmetric
standard model (MSSM) can contain $R$-parity violating interaction
terms :
\beq
{\cal W}_{\rpp} = \lambda_{ijk} \hat L_i \hat L_j \hat
E^c_k + \lambda^\prime_{ijk}\hat L_i \hat Q_j \hat D^c_k +
\lambda^{\prime \prime}_{ijk}\hat U^c_i \hat D^c_j \hat D^c_k
\label{rpv_sup}
\eqn

where, $\hat E^c_i, \hat U^c_i $ and $\hat D^c_i $ are
$i$-th type singlet lepton, up-type and down-type
quark superfields, $\hat L_i$ and $\hat Q_i$ are $SU(2)$ doublet
lepton and quark superfields. The symmetry of the superpotential
requires $\lambda_{ijk} = -\lambda_{jik}$ and
$\lambda^{\prime\prime}_{ijk} = -\lambda^{\prime\prime}_{ikj}$. It
is clear from the Equation \ref {rpv_sup} that the first two terms
violate lepton number, whereas, the last one violate baryon
number. Simultaneous presence of both lepton and baryon number
violating terms in the superpotential will lead to rapid proton
decay. To prevent this we can have either lepton or
baryon number violating terms but not both in the superpotential.
The $\rp$interactions, in general can have
$27 \lambda^\prime $-type and $9$ each of $\lambda$ and
$\lambda^{\prime\prime}$-type of new couplings, which in
general can be complex. The phase of single coupling can be
absorbed in the definition of the sfermion field, but the product
of couplings can have a nontrivial phase. In our analysis we shall
assume that only $ \lambda^\prime_{ijk}$ type of lepton number
violating couplings are present. Furthermore, we will constrain
product of two $\lambda^\prime $ type of couplings at a time
and assume that there are no accidental cancellations and that
only one product dominates at a time.



From the $\rp$superpotential (Equation \ref{rpv_sup}) one can write
down the effective
Lagrangian relevant for our purpose generated by the
exchanging of different sfermions.
\begin{eqnarray}
{\cal L}_{\rpp} &=& {\lambda'_{ijk}\lambda^{'*}_{i'j'k}\over 2 m^2_{\tilde
d^k_R}} \left [ \bar \nu^{i'}_L \gamma^\mu \nu^i_L \bar d^{j'}_L
\gamma_\mu d_L^j + \bar e^{i'}_L \gamma^\mu e^i_L \bar
u_L^{j'}\gamma_\mu u^j_L - \nu^{i'}_L \gamma^\mu e^i_L \bar
d^{j'}_L \gamma_\mu u^j_L -\bar e^{i'}_L \gamma^\mu \nu^i_L \bar
u^{j'}_L \gamma_\mu
d^j_L \right ]\nonumber\\
&-&{\lambda'_{ijk}\lambda^{'*}_{i'jk'}\over 2m^2_{\tilde d^j_L} }
\bar \nu^{i'}_L \gamma^\mu \nu^i_L \bar d^k_R \gamma_\mu d^{k'}_R
-{\lambda'_{ijk}\lambda^{'*}_{i'jk'}\over 2m^2_{\tilde u^j_L}}
\bar e^{i'}_L \gamma^\mu e^i_L \bar d^k_R \gamma_\mu
d^{k'}_R\nonumber\\
&-&{\lambda'_{ijk}\lambda^{'*}_{ij'k'}\over 2m^2_{\tilde e^i_L}}
\bar u^{j'}_{L\beta} \gamma^\mu u^j_{L\alpha} \bar d^k_{R\alpha}
\gamma_\mu d^{k'}_{R\beta}
-{\lambda'_{ijk}\lambda^{'*}_{ij'k'}\over 2m^2_{\tilde \nu^i_L}}
\bar d^{j'}_{L\beta} \gamma^\mu d^j_{L\alpha} \bar d^k_{R\alpha}
\gamma_\mu d^{k'}_{R\beta}, \label{intr}
\end{eqnarray}
In the above $\alpha$ and $\beta$ are color indices.

\subsection{ Constraints from $K^+\to \pi^+ \nu\bar \nu$ process }

In $\rp$MSSM, $ K^+\to \pi^+ \nu\bar \nu$ process receives
two non zero contributions from the exchange of
$\tilde d_R^k$ and $\tilde d_L^j$ squarks. Keeping the SM contribution
one obtains:
\begin{eqnarray}
&&{{\rm Br} (K^+\to \pi^+ \nu \bar \nu)\over {\rm Br} (K^+ \to \pi^0 e^+\nu)}
={r_{K^+}\alpha^2 \over 2 \pi^2 s^4_W |V_{us}|^2}
 \left [ \sum_{\ell=e,\mu,\tau} |\Delta^{SM}_{K} + \Delta^{\rpp}_{K\ell\ell}|^2
 +\sum_{i\neq i'} |\Delta^{\rpp}_{K ii'}|^2 \right ],
\label{kpi_rpv}
 \end{eqnarray}
where, the contribution from the lepton number violating operator
$\hat L_i \hat Q_j \hat D^c_k $ is given by:
\begin{eqnarray}
\Delta^{\rpp}_{K ii'} = {\pi s^2_W\over
\sqrt{2}G_F\alpha} \left [{\lambda'_{i'j2}\lambda^{'*}_{ij1}\over
2m^2_{\tilde d^j_L}} - {\lambda'_{i'1k}\lambda^{'*}_{i2k}\over 2
m^2_{\tilde d^k_R}} \right ].
\label{deltaij}
\end{eqnarray}
There are two contributions arising from the product of two types of
$\lambda^\prime $ couplings, in one case
( $\lambda'_{i'j2}\lambda^{'*}_{ij1}$), the propagator squark coming from
left handed doublet quark superfield $\hat Q_j $, whereas in the other case
( $\lambda'_{i'1k}\lambda^{'*}_{i2k}$), the propagator squark coming from
right handed singlet quark superfield $\hat D^c_k $. In our analysis we take
one combination of couplings to be nonzero at a time by setting all others
to zero. The limits on $\rp$couplings are usually quoted for
$m_{\tilde f} = 100 $ GeV. Following this general practise, through out
our analysis we assume sfermion masses to be degenerate with
$m_{\tilde f} = 100 $ GeV, and limits for higher $m_{\tilde f}$ can be
obtained easily by scaling.

\begin{figure}[hbt]
\centerline{\epsfig{file=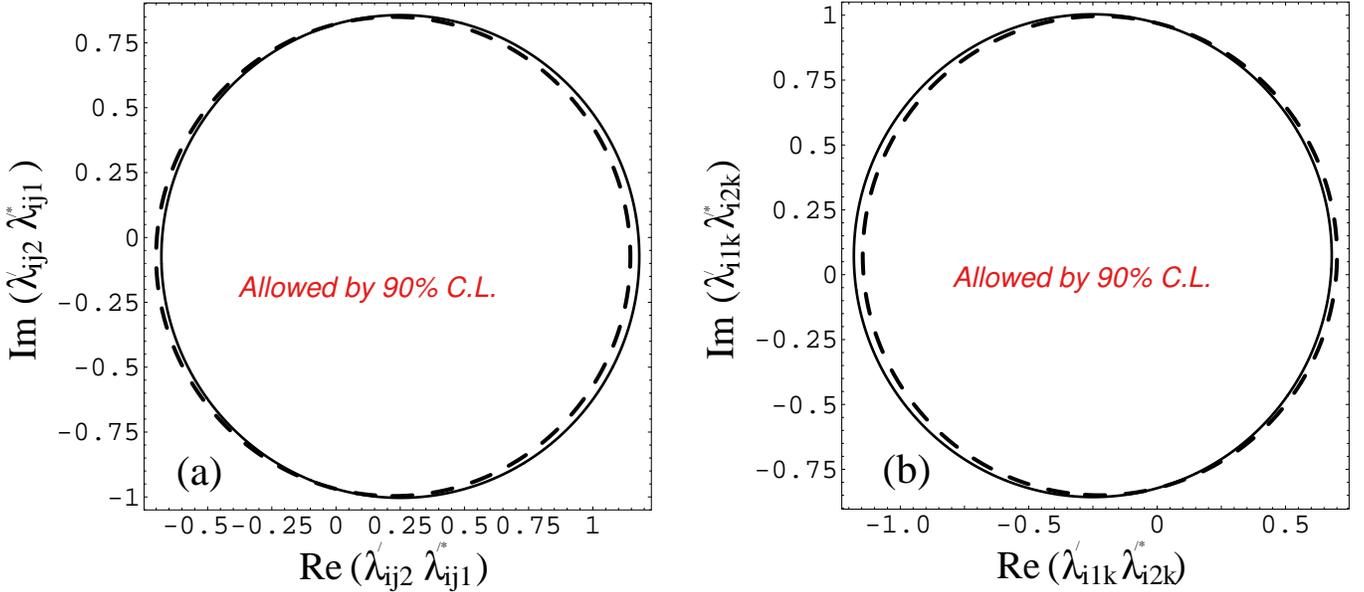,width=\linewidth}}
\vspace*{-13cm}
\caption{ $90\% $ {\rm C.L.} allowed regions for the product of 
$\rp$couplings
(with $i = i^\prime $)(in units of $10^{-5}$) ,
which interfere with the SM from ${\rm Br}~( K^+ \to \pi^+ \nu \bar \nu)$
data. The solid and dashed contours correspond
to $(i= e,\mu)$ and $(i = \tau)$ case respectively.
Figure $(a)$ and
$(b)$ correspond to the first and second combination of $\rp$couplings
in Equation \ref{deltaij}. }
\label{rpv1}
\end{figure}

From Equations \ref{kpi_rpv} and \ref{deltaij} we see that only in
the case of same flavor neutrinos in the final state ($i=i'$),
the $\rp$contributions will interfere with the standard model one.
In Figure \ref{rpv1} $(a) $, the area within the circle represent
the $90 \%$ CL allowed region in the ${\rm Re}
(\lambda^\prime_{ij2}\lambda^{\prime *}_{ij1}) $ - ${\rm Im}
(\lambda^\prime_{ij2}\lambda^{\prime *}_{ij1}) $ (first
combination of
 $\rp$couplings in Equation \ref{deltaij}) plane.
Similarly, the $90\%$ CL allowed regions for the second combination
of $\rp$couplings is shown in the Figure \ref{rpv1} $(b)$ in the
${\rm Re} (\lambda^\prime_{i1k}\lambda^{\prime *}_{i2k}) $ - ${\rm Im}
(\lambda^\prime_{i1k}\lambda^{\prime *}_{i2k}) $ plane.
The solid contour represent $\pi^+ \nu_e\nu_e $ and $\pi^+ \nu_\mu \nu_{\mu}$
(with $i=1,2 $) final state, while the dashed one correspond to
$\pi^+ \nu_\tau \nu_{\tau}$ (with $i = 3$) final state.
The product of $\rp$couplings are in units of $10^{-5}$. The
marginal difference between the solid and dashed contours arises due
the difference in the charm contribution at NLO for $e,\mu$ and $\tau$.
The relative shifts of the bounds between Figure \ref{rpv1} $(a)$ and $(b)$
can traced to the relative sign difference between two combinations of
$\rp$couplings in Equation \ref{deltaij}.

In the scenario, with $i\ne i^\prime $, the $\rp$operators contribute
to the amplitude ${\cal M }(K^+ \to \pi^+ \nu \bar\nu) $ incoherently,
without interfering with the standard model. Moreover, both
combination of $\rp$couplings in Equation \ref{deltaij} will have same
contribution to $ K^+ \to \pi^+ \nu \bar\nu  $ process.
From the experimentally observed branching ratio of
$K^+ \to \pi^+ \nu \bar\nu $ process, we obtains
 $ \mid \lambda^\prime_{i^\prime j 2}\lambda^{\prime}_{ij1}\mid
\leq 0.89 \times 10^{-5} $ at $90\%$ CL for $m_{\tilde d^j_{L}}= 100$ GeV.
The same bound will also apply for the other combination of
$\rp$couplings $ \lambda^\prime_{i^\prime 1 k}\lambda^{\prime *}_{i2k}$ in
Equation \ref{deltaij}.

\begin{table}
\begin{center}
\begin{tabular}{|c|c|c|}
\hline
Couplings & bounds    & source \\
\hline
$\lp_{i1k}\lps_{i2k}$ & $-1.168 \times 10^{-5}
\leq {\rm Re}(\lp_{i1k}\lps_{i2k})\leq 0.67 \times 10^{-5}$ & \\
& $-0.85 \times 10^{-5} \leq {\rm Im} (\lp_{i1k}\lps_{i2k})
\leq 1.0 \times 10^{-5}$ & $K^+\to \pi^+ \nu\bar \nu$\\
\hline
$ \mid \lp_{112} \lp_{111}\mid $ & $ 4.8 \times 10^{-7} $ & $ \Delta M_K$
\cite{kundu_saha}\\
\hline
$ \mid \lp_{122} \lp_{121}\mid $ & $ 4.6 \times 10^{-7} $ & $ \Delta M_K$
\cite{kundu_saha}\\
\hline
$ \lp_{132} \lps_{131} $ & $ -0.67\times 10^{-5} \leq {\rm Re}( \lp_{132} \lps_{131})
\leq 1.17 \times 10^{-5} $ & \\
     &  $ -1.0\times 10^{-5} \leq {\rm Im}(  \lp_{132} \lps_{131})   \leq 0.85
\times 10^{-5} $ &  $ K^+ \to \pi^+ \nu \bar \nu $\\
\hline
$ \mid \lp_{212}\lp_{211}\mid $ & $ 4.8 \times 10^{-7}$ & $
\Delta M_K $\cite{kundu_saha}\\
\hline
$ \mid \lp_{222}\lp_{221}\mid $ & $ 4.6 \times 10^{-7}$ & $
\Delta M_K $\cite{kundu_saha}\\
\hline
$ \lp_{232} \lps_{231} $ & $ -3.719 \times 10^{-6} \leq {\rm Re}( \lp_{232} \lps_{231})
\leq 1.14 \times 10^{-6} $ &  $ K_L \to \mu^+ \mu^- $\\
     &  $ -1.0\times 10^{-5} \leq {\rm Im}(  \lp_{232} \lps_{231})   \leq 0.85
\times 10^{-5} $ &  $ K^+ \to \pi^+ \nu \bar \nu $\\
\hline
$ \mid \lp_{312} \lp_{311}\mid $ & $ 4.8 \times 10^{-7} $ & $ \Delta M_K$
\cite{kundu_saha}\\
\hline
$\lp_{3j2}\lps_{3j1}$ & $-0.67\times 10^{-5} \leq {\rm Re}( \lp_{3j2} \lps_{3j1})
\leq 1.168 \times 10^{-5} $ & \\
$(j = 2,3)$     &  $ -1.0\times 10^{-5} \leq {\rm Im}(  \lp_{3j2} \lps_{3j1})   \leq 0.85
\times 10^{-5} $ &  $ K^+ \to \pi^+ \nu \bar \nu $\\
\hline
$\mid \lp_{11k}\lp_{22k}\mid $ & $ 4 \times 10^{-7}$ &
$ \mu {\rm Ti} \to e {\rm Ti} $ \cite{huitu}\\
\hline
$ \mid \lp_{21k}\lp_{12k} \mid $ & $  4.3 \times 10^{-7}$ &
$ \mu {\rm Ti} \to e {\rm Ti} $ \cite{huitu}\\
\hline
$ \mid \lp_{11k}\lp_{32k}\mid $ & $ 0.89 \times 10^{-5} $
& $ K^+ \to \pi^+ \nu \bar \nu $ \\
\hline
$ \mid \lp_{21k}\lp_{32k}\mid $ & $ 0.89 \times 10^{-5} $
& $ K^+ \to \pi^+ \nu \bar \nu $ \\
\hline
$ \mid \lp_{31k}\lp_{12k}\mid $ & $ 0.89 \times 10^{-5} $
& $ K^+ \to \pi^+ \nu \bar \nu $ \\
\hline
$ \mid \lp_{31k}\lp_{22k}\mid $ & $ 0.89 \times 10^{-5} $
& $ K^+ \to \pi^+ \nu \bar \nu $ \\
\hline
$\mid \lp_{i^\prime j 2}\lp_{ij1}\mid $ $\clubsuit $ & $ 0.89 \times 10^{-5} $
& $ K^+ \to \pi^+ \nu \bar \nu $ \\
\hline
\end{tabular}
\caption{ Current relevant upper bounds on the values of products of
two $\rp$couplings. Bounds corresponding to Ref. \cite{kundu_saha} and
Ref.\cite{huitu} for certain combinations are stronger than the one obtained
in here.
$\clubsuit$ $ K_L \to \mu e $ process
put stronger limits $( 8 \times 10^{-7})$ on the combinations:
$\mid \lp_{212}\lp_{111}\mid , \mid \lp_{222}\lp_{121} \mid $
and $ \mid \lp_{232}\lp_{131} \mid $ \cite{barbieri}.} 
\end{center}
\label{rpv_bound}
\end{table}

\noindent {\bf Constraints from $K_L \to \mu^+ \mu^-$ process }

The $K_L \to \mu^+ \mu^- $ process bounds the real part of the
product $\lambda^\prime_{2i1}\lambda^{\prime *}_{2i2}$.
The expression of the branching ratio
${\rm Br} ( K_L \to \mu^+ \mu^-)_{\rm SD} $ has been
given in Equation (\ref{klmm}).
The $\rp$contribution can be easily included by
replacing $\Delta^Z_{K_L}$ in Equation (\ref{klmm1}) by
$-(\pi s^2_W/\sqrt{2}G_F\alpha) \lambda'_{2j1}\lambda^{'*}_{2j2}
/2m^2_{\tilde u^j_L}$. Using the $90\%$ CL upper bound on the
${\rm Br} ( K_L \to \mu^+ \mu^-)_{\rm SD} \leq 2.5 \times 10^{-9}$ (as
mentioned before), we obtain
\beq
-3.719 \times 10^{-6} < {\rm Re} (\lambda^\prime_{2j1}\lambda^{\prime *}_{2j2})
< 1.14 \times 10^{-6}.
\eqn
for $m_{\tilde u^j_L} = 100 $ GeV.

\noindent { \bf Constraints from $K^+ \to \pi^+ \mu^+ e^- $ process }

This lepton number violating process has no contribution from the
standard model. However, in $\rp$model,
$\lambda^\prime_{1j1}\lambda^{\prime *}_{2j2}$ coupling can induce such
decay process mediated by virtual $\tilde u_L$.
The branching ratio is given by
\beq
\frac{ {\rm Br}(K^+ \to \pi^+ \mu^+ e^-)}{{\rm Br} (K^+ \to \pi^0 e^+ \bar\nu)}
&=& \frac{r_{K^+} \alpha^2}{2 \pi^2 s^4_W \mid V_{us}\mid^2}
\mid \Delta^{\rpp}_{K^+}\mid^2, \\
\Delta^{\rpp}_{K^+} &=&  \frac{ \pi s^2_W }{ \sqrt{2} G_F \alpha }
\left ( \frac{ \lambda^\prime_{1j1} \lambda^{\prime *}_{2j2} }
{ 2 m^2_{ \tilde u^j_L }}\right )
\eqn
Experimentally we only have an upper bound on the branching ratio,
${\rm Br}(K^+ \to \pi^+ \mu^+ e^-) < 2.8 \times 10^{-11}$ at $90\%$ CL.
Using this bound we obtain at $90\%$ CL
\beq
\mid \lambda^\prime_{1j1}\lambda^{\prime *}_{2j2} \mid
\leq 2.684 \times 10^{-6}
\eqn
for $m_{\tilde u^j_L} = 100 $ GeV.
However, in Ref.\cite{barbieri} authors have obtained stronger bound of
$\mid \lambda^\prime_{1j1}\lambda^{\prime *}_{2j2} \mid
\leq 8.0\times 10^{-7}$ from $K_L \to \mu e $ process.

We now display in Table I the best
current upper bounds on the product of two
$\rp$couplings with the processes which provides the limit.
We present bounds, obtained by us from
$K^+ \to \pi^+ \nu \bar \nu$ as well as some earlier
bounds from other processes like $\Delta M_K $ and  $\mu \to e $ conversion
in the nuclei which are some times stronger.

From our analysis, it is clear that the bounds obtained on the
different combinations of $\rp$couplings from $K^+ \to \pi^+ \nu
\bar \nu$ process is of the order of $10^{-5}$. In the case where
the pair of $\rp$couplings which interfere with the standard model
we obtain bounds on real and imaginary part of the pair separately
from Figure \ref{rpv1}, while in the non-interfering case, we
obtain bound on the magnitude of the pair of $\rp$couplings
involved. Note that earlier bounds on the above combination were
$\sim {\cal O}(10^{-3})$ allowing significant room for enhancement
of the process $K^+ \to \pi^+ \nu \bar \nu $. The other process
$K_L \to \mu^+ \mu^-$ puts a limit on ${\rm Re}
(\lp_{232}\lps_{231})$, which is of the order of $10^{-6}$. The
$\Delta M_K$ and $\mu \to e $ conversion in the nuclei provide
bounds of the order of $10^{-7}$ on some combinations of
$\rp$couplings which are otherwise weakly constrained by our
analysis. \footnote {Note that we do not use
$\epsilon^\prime/\epsilon $ to put bounds on products of $\lp$
couplings because the theoretical expression for this quantity
involves $\rp$scalar couplings, involving more model dependence
\cite{abel}.}

\noindent {\bf Prediction for $ K_L \to \pi^0 \nu \bar \nu $ process }

In the presence of $L$ number violating $\lp_{ijk} L_iQ_j
D^c_k$ operator, several products of two $\rp$couplings can
contribute to the $CP$ violating process $K_L \to \pi^0 \nu
\bar\nu$. We have
\begin{eqnarray}
{{\rm Br} (K_L\to \pi^0 \nu \bar \nu)\over {\rm Br}(K^+ \to \pi^0 e^+\nu)}
&=& \kappa \left [\sum_{\ell=e,\mu,\tau}\mid {\rm Im}(\Delta^{SM}_{K}
+ \Delta^{\rpp}_{K_L \ell\ell})\mid^2
 +\sum_{i\neq i'}
\mid \Delta^{\rpp}_{K_L ii'} \mid^2 \right ]\nonumber\\
\kappa &=& r_{K_L}{\tau_{K_L}\over \tau_{K^+}}{\alpha^2 \over 2 \pi^2
s^4_W |V_{us}|^2}
\eqn

\begin{eqnarray}
\Delta^{\rp}_{K_Lii'} = {\pi s^2_W\over \sqrt{2}G_F\alpha}\left [
{\lambda'_{i'j1}\lambda^{'*}_{ij2}\over 2m^2_{\tilde d_L^j}} -
{\lambda'_{i'2k}\lambda^{'*}_{i1k}\over 2 m^2_{\tilde d^k_R}}
-{\lambda'_{i'j2}\lambda^{'*}_{ij1}\over 2m^2_{\tilde d_L^j}} +
{\lambda'_{i'1k}\lambda^{'*}_{i2k}\over 2 m^2_{\tilde d^k_R}} \right ].
\label{klpinn}
\end{eqnarray}

One notes that for $i=i'$, the decays are $CP$ violating, but for
$i\neq i'$, the decays are not necessarily $CP$ violating which is
very different from the SM\cite{mu}.

As we see from the Equation \ref{klpinn}, that several combinations
of two $\rp$couplings are involved in this case. In our numerical calculation
we consider each combination of $\rp$couplings one by one.
For $i = i^\prime $ these $\rp$contributions interfere with the standard
model one, while for
$i \ne i^\prime $, these new operators create neutrino pair which are
not $CP$ eigenstate.
We will treat these two cases separately. First we assume $i = i^\prime$,
in this
case the branching ratio ${\rm Br}(K_L\to \pi^0 \nu \bar \nu)$ can reach
as large as $2.0 \times 10^{-9}$ for the allowed values of
$\lp_{131}\lps_{132}, \lp_{232}\lps_{231},\lp_{322}\lps_{321},\lp_{332}
\lps_{331} $ couplings.
In fact, we find that for all the relevant pairs of
$\rp$couplings (with $i = i^\prime $),
whose bound is $\sim {\cal O}(10^{-5})$, the
maximum value of the branching ratio is $2.0 \times 10^{-9}$, which is
almost two order of magnitude larger than the standard model prediction:
$2.93^{+0.84}_{-0.67}\times 10^{-11}$. Experimentally we only have an
upper limit for this branching ratio, which is $5.9 \times 10^{-7}$ at
$90\%$ CL. In the second scenario, where $i \ne i^\prime $, $K_L \to \mu e$
and $\mu {\rm Ti} \to e {\rm Ti}$ set bounds of the order of
${\cal O}(10^{-7})$ on the magnitude
of following combination of $ \rp$couplings: $\mid \lp_{11k}\lp_{22k}\mid,
\mid\lp_{21k}\lp_{12k}\mid, \mid\lp_{212}\lp_{111}\mid,
\mid \lp_{222}\lp_{121}\mid $ and  $ \mid \lp_{232}\lp_{131}\mid $. As can be
seen from the Table I that the bound on the magnitude of
other combinations of $\rp$couplings are $0.89 \times 10^{-5} $ obtained from
$K^+ \to \pi^+ \nu \bar \nu $. We find that the branching ratio can reach up
to $1.39 \times 10^{-9}$ for the pair of $\rp$couplings in
Equation \ref{klpinn} whose magnitude satisfy the above mentioned limit.

\section{Conclusions}

In this paper we have examined the effects of new physics originated from
two different kind of models on several rare flavor changing processes
involving $K$ meson: $K^+ \to \pi^+ \nu \bar\nu$,$ K_L \to \mu^+ \mu^-$,
$\epsilon^\prime/\epsilon $, $ K_L \to \pi^0 e^+e^-$,
$K_L \to \pi^0 \nu \bar \nu$.
In the first model, in addition to the SM quarks we also have
one extra down-type singlet quark. The presence of such an additional
singlet quark leads to a new off-diagonal $Z$ mediated FCNC coupling
$U_{ij}$, between the SM quarks of flavor $i$ and $j$. In this paper we
have considered off-diagonal $Z$ coupling between first two generation
down type SM quarks, denoted by a complex parameter $U_{sd}$. We then
obtained $90\%$ C.L. bound on this mixing parameter using known
experimental data on $K^+ \to \pi^+ \nu \bar \nu$, $K_L \to \mu^+ \mu^-$
and $\epsilon^\prime/\epsilon$. It turned out that the allowed parameter
space of $U_{sd}$ from $K^+ \to \pi^+\nu \bar \nu$ process is severely
constrained from $\epsilon^\prime/\epsilon$ ( imaginary part of $U_{sd}$):
$ -1.91\times 10^{-6} \leq {\rm Im} (U_{sd}) \leq 3.72 \times 10^{-6}$
, while  the $(K_L \to \mu^+ \mu^-)_{\rm SD}$ constrains the real part of
$U_{sd}$: $ -1.12 \times 10^{-5} \leq {\rm Re} (U_{sd})
\leq 3.45\times 10^{-6}$. We did not find any significant deviation from
the SM prediction of $K^0 -\bar K^0$ oscillation.
The value of ${\rm Br} (K^+\to \pi^+ \nu \bar \nu)$ can reach up to
$1.5\times 10^{-10}$ at the edge of the allowed parameter space of $U_{sd}$.
Moreover, we can also reproduce the central value of experimentally
measured ${\rm Br}(K^+ \to \pi^+ \nu \bar \nu)$. We have also studied
other $CP$ violating kaon processes, $K_L \to \pi^0 \nu \bar \nu $ and
$K_L \to \pi^0 e^+ e^- $ in the allowed parameter
space of $U_{sd}$. The value of ${\rm Br}(K_L \to \pi^0 \nu \bar \nu)$ can
reach as high as $10\times 10^{-11}$ at the edge of the allowed parameter
space. At present from experiment we have an upper limit on this branching
ratio $5.9 \times 10^{-7}$ at $90 \%$ CL \cite{prot_life}.
We have found that the process $K_L \to \pi^0 e^+ e^-$ is very weakly dependent
on the new physics parameter $U_{sd}$, because of the fact that the dominant
contribution to this decay amplitude arises from the mixing of $K_L$ and
$K_S$, followed by $K_S \to \pi^0 e^+ e^-$ decay.  

The second model, we have considered is the $\rp$MSSM.
We have computed the bounds on the product of two $\rp$couplings of the type
$\lp \lp$ using $K^+ \to \pi^+ \nu \bar \nu$, $K_L \to \mu^+ \mu^-$ and
$K^+ \to \pi^+ \mu^+ e^-$ processes. We have assumed that the product of
two $\rp$couplings are in general complex and all the sfermion masses are
degenerate with mass of $100$~GeV in order to compare with earlier bounds
obtained in literature. One can obtain bounds for any other sfermion mass by
scaling. In deriving the bounds full standard model amplitudes have been
taken
into account. One should note that in several cases, our bounds are
complement to the bounds obtained from other processes like $\Delta M_K,
K_L \to \mu e $ and $\mu {\rm Ti} \to e {\rm Ti}$. We have found that
processes like $K^+ \to \pi^+ \nu\bar\nu $ $ K_L \to \pi^0 \nu \bar \nu $ 
can be significantly enhanced compared to their standard model predictions.
The constraints on the product of $\rp$couplings 
$\lambda^\prime_{131}\lambda^{*\prime}_{132}$ from  the 
decay mode $K_L \to \pi^0 e^+ e^-$ is $\sim {\cal O}(10^{-4})$, which is 
one order of magnitude weaker than the bound obtained from 
$K^+ \to \pi^+ \nu \bar\nu$. As we have explained before, the dominant 
contribution to the $K_L \to \pi^0 e^+ e^-$ process arises from the
mixing between $K_L$ and $K_S$. After taking into account this mixing
contribution, the standard model prediction for the
${\rm Br} (K_L \to \pi^0 e^+e^-) = (3.2^{+1.2}_{-0.8})\times 10^{-11}$ 
\cite{buchalla_1}, whereas
the experimental upper bound in $5.1\times 10^{-10}$ at $90\%$CL, leaving 
a very small room for the new physics contribution. 

\begin{flushleft}
\begin{large}
{\bf Acknowledgments}
\end{large}
\end{flushleft}
This work was supported in part by US DOE contract numbers
DE-FG03-96ER40969, and supported in part by NSC.

\def\pr#1,#2 #3 { {Phys.~Rev.}        ~{\bf #1},  #2 (19#3) }
\def\prd#1,#2 #3{ { Phys.~Rev.}       ~{D \bf #1}, #2 (19#3) }
\def\pprd#1,#2 #3{ { Phys.~Rev.}      ~{D \bf #1}, #2 (20#3) }
\def\prl#1,#2 #3{ { Phys.~Rev.~Lett.}  ~{\bf #1},  #2 (19#3) }
\def\pprl#1,#2 #3{ {Phys. Rev. Lett.}   {\bf #1},  #2 (20#3)}
\def\plb#1,#2 #3{ { Phys.~Lett.}       ~{\bf B#1}, #2 (19#3) }
\def\pplb#1,#2 #3{ {Phys. Lett.}        {\bf B#1}, #2 (20#3)}
\def\npb#1,#2 #3{ { Nucl.~Phys.}       ~{\bf B#1}, #2 (19#3) }
\def\pnpb#1,#2 #3{ {Nucl. Phys.}        {\bf B#1}, #2 (20#3)}
\def\prp#1,#2 #3{ { Phys.~Rep.}       ~{\bf #1},  #2 (19#3) }
\def\zpc#1,#2 #3{ { Z.~Phys.}          ~{\bf C#1}, #2 (19#3) }
\def\epj#1,#2 #3{ { Eur.~Phys.~J.}     ~{\bf C#1}, #2 (19#3) }
\def\eepj#1,#2 #3{ { Eur.~Phys.~J.}     ~{\bf C#1},#2 (20#3) }
\def\mpl#1,#2 #3{ { Mod.~Phys.~Lett.}  ~{\bf A#1}, #2 (19#3) }
\def\ijmp#1,#2 #3{{ Int.~J.~Mod.~Phys.}~{\bf A#1}, #2 (19#3) }
\def\ptp#1,#2 #3{ { Prog.~Theor.~Phys.}~{\bf #1},  #2 (19#3) }
\def\jhep#1,#2 #3{ {J. High Energy Phys.} {\bf #1}, #2 (19#3)}
\def\pjhep#1,#2 #3{ {J. High Energy Phys.} {\bf #1}, #2 (20#3)}

\end{document}